%% file: star.tex
\documentclass{article}

\usepackage{epsfig}
\usepackage{latexsym}
\usepackage{amsmath}
\usepackage{amssymb}

\newtheorem{theorem}{Theorem}[section]

\newtheorem{lemma}[theorem]{Lemma}

\newcommand{\PP}{\mathbb P}
\newcommand{\EE}{{\mathbb E}}

\newcommand{\old}[1]{{}}

\title{The Bayesian `star paradox' persists  for long finite sequences}
\author{Mike Steel and Frederick A. Matsen\\
Allan Wilson Centre for Molecular Ecology and Evolution \\ 
\\ \\
Corresponding author:\\
Mike Steel \\
Biomathematics Research Centre\\ 
Department of Mathematics and Statistics\\
University of Canterbury\\
Private Bag 4800\\
Christchurch, New Zealand\\
Phone: +64-3-364-2987 ext. 7688\\
Fax: +64-3-364-2587\\
Email: M.Steel@math.canterbury.ac.nz
}

\begin{document}

\maketitle

{\noindent Keywords: phylogenetic trees, Bayesian statistics, star trees}

\vspace{-4pt}
{\noindent Running head: The star paradox persists}

\newpage

\begin{abstract}
The `star paradox' in phylogenetics is the tendency for a particular resolved tree to be sometimes strongly supported even when the data is generated by an unresolved (`star') tree. There have been contrary claims as to whether this phenomenon persists when very long sequences are considered. This note settles one aspect of this debate by proving mathematically that there is always a chance that a resolved tree could be strongly supported, even as the length of the sequences becomes very large. 
\end{abstract}

\section{Introduction}

Two recent papers (Yang and Rannala 2005; Lewis, Holder and
Holsinger 2005) highlighted a phenomenon that occurs when sequences
evolve on a tree that contains a polytomy - in particular a
three-taxon unresolved rooted tree.   As longer sequences are analysed
using a Bayesian approach, the posterior probability of the trees that
give the different resolutions of the polytomy do not converge on
relatively equal probabilities - rather a given
resolution can sometimes have a posterior probability close to one.  In
response Kolaczkowski and Thornton (2006) investigated this phenomena
further, providing some interesting simulation results, and offering
an argument that seems to suggest that for very long sequences the
tendency to sometimes infer strongly supported resolutions suggested by
the earlier papers would disappear with sufficiently long sequences.
As part of their case the authors use the expected site frequency
patterns to simulate the case of infinite length sequences, concluding that
``with infinite length data, posterior probabilities give equal
support for all resolved trees, and the rate of false inferences falls
to zero."  Of course these findings concern sequences that are
effectively infinite, and, as is well known in statistics, the limit
of a function of random variables (in this case site pattern
frequencies for the first $n$ sites) does not necessarily equate with
the function of the limit of the random variables. Accordingly
Kolaczkowski and Thornton
offer this appropriate cautionary qualification of their findings:

``Analysis of ideal data sets does not indicate what will happen when
very large data sets with some stochastic error are analyzed, but it does
show that when infinite data are generated on a star tree, posterior
probabilities are predictable, equally supporting each possible resolved
tree."

Yang and Rannala (2005) had attempted to simulate the large sample
posterior distribution, but ran into numerical problems and commented
that it was ``unclear'' what the  limiting distribution on
posterior probabilities was as $n$ became large.

In particular, all of the aforementioned papers have left open an
interesting statistical question, which this short note formally answers
- namely, does the Bayesian posterior probability of the three
resolutions of a star tree on three taxa converge to 1/3 as the sequence
length tends to infinity? That is, does the distribution on posterior
probabilities for `very long sequences' converge on the distribution for
infinite length sequences?  We show that for most reasonable priors it
does not. Thus the `star paradox' does not disappear as the
sequences get longer.

As noted by (Yang and Rannala 2005; Lewis, Holder and Holsinger 2005) one can demonstrate such phenomena more
easily for related simpler processes such as coin tossing (particularly if one imposes a particular prior). Here we
avoid this simplification to avoid the criticism that such results do
not rigorously establish corresponding phenomena in the phylogenetic setting,
which in contrast to coin tossing involves considering a parameter
space of dimension greater than 1. We also frame our main result so that it applies to a fairly general class of priors.
Note also that it is not the purpose of this short note to add to
the on-going debate concerning  the implications of this `paradox' for Bayesian phylogenetic
analysis, we merely demonstrate its existence. Some further comments and
earlier references on the phenomenon have been described in the recent review
paper by Alfaro and Holder 2006 (pp. 35-36). 
\begin{figure}[h]
\begin{center} \label{starfig}
\resizebox{12cm}{!}{
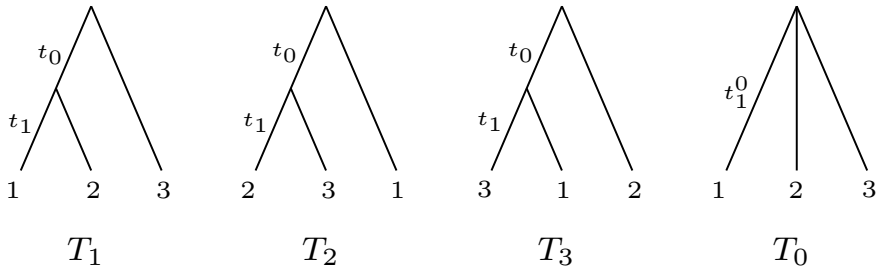
}
\caption{The three resolved rooted phylogenetic trees on three taxa $T_1, T_2, T_3$, and the unresolved `star' tree on which the sequences are generated $T_0$.}
\end{center}
\label{overview}
\end{figure}

\section{Analysis of the star tree paradox for three taxa}
On tree $T_1$ (in Fig. 1) let $p_i = p_i(t_0, t_1)$, $i = 0,1,2,3$ denote the probabilities of the four
site patterns ($xxx, xxy, yxx, xyx$, respectively) under the simple $2$--state
symmetric Markov process (the argument extends to more general models,
but it suffices to demonstrate the phenomena for this simple model). 
From Eqn. (2) of (Yang and Rannala 2005) we have
$$p_0(t_0, t_1) = \frac{1}{4}(1+e^{-4t_1} + 2e^{-4(t_0+t_1)}),$$
$$p_1(t_0, t_1) = \frac{1}{4}(1+e^{-4t_1} -2e^{-4(t_0+t_1)}),$$
and 
$$p_2(t_0, t_1) = p_3(t_0, t_1) = \frac{1}{4}(1-e^{-4t_1}).$$
It follows by elementary algebra  that for $i=2,3$,
\begin{equation}
\label{ineq1}
\frac{p_1(t_0, t_1)}{p_i(t_0, t_1)} \geq 1+ 2e^{-4t_1}(1-e^{-4t_0}),
\end{equation}
 and thus $p_1(t_0, t_1) \geq p_i(t_0, t_1)$ with strict inequality unless 
$t_0=0$ (or in the limit as $t_1$ tends to infinity).  

To allow maximal generality we make only minimal neutral assumptions about the
prior distribution on  trees and branch lengths. Namely  we assume that the three resolved trees on three leaves (trees $T_1, T_2, T_3$ in Fig. 1) have equal prior probability $\frac{1}{3}$ and that the prior distribution on branch lengths
$t_0, t_1$ is the same for each tree,  and  has a continuous joint
probability density function that is everywhere non-zero. This condition
applies for example to the exponential and gamma priors discussed by Yang and Rannala
(2005).  Any prior that satisfies these conditions we call {\em reasonable}. Note
that we do not require that $t_0$ and $t_1$ be independent.  

Let ${\bf n} = (n_0, n_1, n_2, n_3)$ be the counts of the different types of
site patterns (corresponding to the same patterns as for the $p_i$'s). Thus $n= \sum_{i=0}^3 n_i$ is the total number of sites (i.e. the length of the sequences). 
Given a prior distribution on $(t_0, t_1)$ for the branch lengths of $T_i$ (for $i=1,2,3$) 
let $\PP[T_i|{\bf n}]$ be the posterior probability of tree $T_i$ given the site pattern counts ${\bf n}$.
Now suppose the $n$ sites are generated on a star tree $T_0$ with positive branch lengths.  We are interested in whether  the posterior probability $\PP[T_i|{\bf n}]$ could be close to 1 or whether the chance of generating data with this property goes to zero as the sequence length gets very large. We show that in fact the latter possibility is ruled out by our main result, namely the following:

\begin{theorem}
\label{main}
Consider sequences of length $n$ generated by a star tree $T_0$ on three
taxa with strictly positive edge length $t_1^0$ and let ${\bf n}$ be
the resulting data (in terms of site pattern counts). Consider any prior
on the three resolved trees $(T_1, T_2, T_3)$ and their branch lengths that is reasonable (as defined above).  For any $\epsilon >0$, and each resolved tree $T_i$
($i=1,2,3$),
the probability that ${\bf n}$ has the property that $$\PP(T_i|{\bf n}) >
1-\epsilon$$  does not converge to $0$ as $n$ tends to infinity. 
\end{theorem}
{\em Proof of Theorem~\ref{main}}
Consider the star tree $T_0$ with given branch lengths
$t_1^0$ (as in Fig. 1). Let $(q_0, q_1, q_2, q_3)$ denote the probability of the four types of site patterns generated by $T_0$ with these branch lengths. Note that $q_1=q_2=q_3$ and so $q_0 = 1-3q_1$).
Suppose we generate $n$ sites on this tree, and let $n_0, n_1, n_2, n_3$ be the counts of the different types of
site patterns (corresponding to the $p_i$'s). Let $\Delta_0: = \frac{n_0-q_0n}{\sqrt{n}}$ and 
for $i=1,2,3$ let $$\Delta_i:= \frac{n_i - \frac{1}{3}(n-n_0)}{\sqrt{n}}.$$
For a  constant $c>1$, let $F_c$ denote the event: $$F_c:  \Delta_2,
\Delta_3 \in [-2c, -c] \mbox { and } \Delta_0 \in [-c,c].$$ Notice that
$F_c$ implies $\Delta_1 \in  [2c, 4c]$ since
$\Delta_1+\Delta_2+\Delta_3=0$. By standard stochastic arguments (based
on the asymptotic approximation of the multinomial distribution by the
multinormal distribution) event $F_c$ has probability at least some value
$\delta'=\delta'(c)>0$ for all $n$ sufficiently large (relative to
$c$). 

Given the data ${\bf n}=(n_0, n_1, n_2, n_3)$ the assumption of equality
of priors across $T_1, T_2$ and $T_3$ implies that
\begin{equation}
\label{identities1}
\PP(n_0, n_1, n_2, n_3|T_2, t_0, t_1) = \PP(n_0, n_2, n_3, n_1|T_1,
t_0, t_1),
\end{equation}
 and
\begin{equation}
\label{identities2}
\PP(n_0, n_1, n_2, n_3|T_3, t_0, t_1) = \PP(n_0, n_3, n_1, n_2|T_1,
t_0, t_1). 
\end{equation}

Now, as $(t_0, t_1)$ are random variables with some prior density, when
we view $p_0, p_1, p_2, p_3$ as random variables by virtue of their
dependence on $(t_0, t_1)$, we will write them as $P_0, P_1, P_2,
P_3$ (note that Yang and Rannala (2005) use $P_i$ differently). With this notation, the posterior probability of $T_1$ conditional on ${\bf n}$ can be
written as
$$\PP(T_1|{\bf n})= p({\bf n})^{-1}\times\EE_1[P_0^{n_0}P_1^{n_1}P_2^{n_2}P_3^{n_3}]$$
where $ p({\bf n})$ is the posterior probability of ${\bf n}$ 
and $\EE_1$ denotes expectation with respect to the prior for
$t_0, t_1$ on $T_1$. 
Moreover since $P_2=P_3$, we can write this as
$\PP(T_1|{\bf n})=  p({\bf n})^{-1}
\times\EE_1[P_0^{n_0}P_1^{n_1}P_2^{n_2+n_3}].$ By (\ref{identities1}) and (\ref{identities2})  we have
$$\PP(T_2|{\bf n}) = p({\bf n})^{-1} \times \EE_1[P_0^{n_0}P_1^{n_2}P_2^{n_1+n_3}]; \mbox{ and } \PP(T_3|{\bf n}) = p({\bf n})^{-1} \times \EE_1[P_0^{n_0}P_1^{n_3}P_2^{n_1+n_2}]$$
where again expectation is taken with respect to the prior for $t_0,
t_1$ on $T_1$. Consequently, 
\begin{equation}
\label{eqfrac}
\frac{\PP(T_1|{\bf n})}{\PP(T_2|{\bf n})} = \frac{\EE_1[X]}{\EE_1[Y]},
\end{equation}
where $$X= P_0^{n_0}P_1^{n_1}P_2^{n_2+n_3} \mbox{ and }
Y=P_0^{n_0}P_1^{n_2}P_2^{n_1+n_3}.$$ As will be shown later, it
suffices to demonstrate that the ratio in (\ref{eqfrac}) 
can be large with nonvanishing
probability in order to obtain the conclusion of the theorem. In order
to do so we use the following lemma, whose proof is provided in the Appendix.

\begin{lemma}
\label{lem3}
Let $X,Y$ be non-negative continuous random variables, dependent on a third random variable $\Lambda$ that takes values in an interval $I=[a,b]$.  Suppose that for some interval $I_0$ strictly inside $I$, and $I_1=I-I_0$ the following inequality holds:
\begin{equation}
\label{eqin}
\EE[Y|\Lambda \in I_0] \geq \EE[Y|\Lambda \in I_1],
\end{equation}  
and that for some constant $B>0$, and all $\lambda \in I_0$, 
\begin{equation}
\label{eqin2}
\frac{\EE[X|\Lambda=\lambda]}{\EE[Y|\Lambda=\lambda]} \geq B. 
\end{equation}  Then,
$\frac{\EE[X]}{\EE[Y]} \geq B\cdot \PP(\Lambda \in I_0)$.
\end{lemma}
\noindent 
To apply this lemma,
select a value $s>0$ so that $\frac{1}{4} + s < q_0 < 1-s$, and let 
$I_0 = [q_0-s, q_0+s]$. Then let
$I = [\frac{1}{4}, 1]$ and $I_1 = I - I_0$. 

{\em Claim:} For $n$ sufficiently large, and conditional on the data
${\bf n} = (n_0, n_1, n_2, n_3)$ satisfying $F_c$:
\begin{itemize}
\item[(i)]  $\EE_1[Y|P_0 \in I_0] \geq \EE_1[Y|P_0 \in I_1]$
\item[(ii)] For all $p_0 \in I_0$,  $\frac{\EE_1[X|P_0=p_0]}{\EE_1[Y|P_0=p_0]}
  \geq 6c^2$.
\end{itemize}
The proofs of these two claims is given in the Appendix.

Applying Lemma~\ref{lem3} to the Claims (i) and (ii) we deduce that
conditional on ${\bf n}$ satisfying $F_c$ and
$n$ being sufficiently large, 
\begin{equation}
\label{cbound}
\frac{\EE_1[X]}{\EE_1[Y]} \geq 6c^2 \cdot \PP(P_0 \in I_0).
\end{equation}

\noindent
Select $c > \frac{1}{\sqrt{3\epsilon \PP(P_0 \in I_0)}}$ (this is
finite by the assumption that the prior on $(t_0, t_1)$ is everywhere
non-zero).   As stated before, the probability that ${\bf n}$ satisfies $F_c$ is at least $\delta'=\delta'(c)>0$  for $n$ sufficiently large.
Then, $6c^2 \cdot \PP(P_0 \in I_0) > \frac{2}{\epsilon}$
and so by (\ref{cbound}), $\frac{\PP(T_1|{\bf n})}{\PP(T_2|{\bf n})} = \frac{\EE_1[X]}{\EE_1[Y]} > \frac{2}{\epsilon}.$
Similarly, $\frac{\PP(T_1|{\bf n})}{\PP(T_3|{\bf n})}  > \frac{2}{\epsilon}.$
Now, since $\PP(T_1|{\bf n})+\PP(T_2|{\bf n})+\PP(T_3|{\bf n})=1$ it now
follows that, for $n$ sufficiently large, and conditional an event of
probability at least $\delta'>0$, that $\PP(T_1|{\bf n})> 1-\epsilon$ as claimed.  This completes the proof. 
\hfill$\Box$

\newpage
\subsection{Concluding remarks}

One feature of the argument we have provided is that it does not require
stipulating in advance a particular prior on the branch lengths -- that
is, the result is somewhat generic as it imposes relatively few
conditions. Moreover,
 the requirement that the prior on $(t_0, t_1)$ be everywhere
non-zero could be weakened to simply being non-zero in a neighborhood of $(0,
t_1^0)$ (thereby allowing, for example,  a uniform distribution on
bounded range). 

A interesting open question in the spirit of this paper is to
explicitly calculate the limit of the posterior density $f(P_1, P_2,
P_3)$ described in (Yang and Rannala 2005).

\subsection{Acknowledgments} MS thanks Ziheng Yang for suggesting the
problem of computing the limiting distribution of posterior
probabilities for 3-taxon trees. This work is funded by the 
\emph{Allan Wilson Centre for Molecular Ecology and Evolution}.

\section*{References}
\noindent
Alfaro ME, Holder MT. 2006. The posterior and prior in Bayesian
phylogenetics. Annu. Rev. Evol. Syst. 37: 19-42.

\noindent
Kolaczkowski B, Thornton JW. 2006. 
Is there a star tree paradox? Mol. Biol. Evol. 23: 1819--1823.

\noindent
Lewis PO, Holder MT, Holsinger KE. 2005. Polytomies and Bayesian
phylogenetic inference. Syst. Biol. 54 (2): 241-253.

\noindent
Yang Z, Rannala B. 2005. Branch-length prior influences Bayesian
posterior probability of phylogeny. 
Syst. Biol. 54 (3): 455-470.

\newpage

\section{Appendix: Proof of Lemma ~\ref{lem3} and Claims (i), (ii)}

{\em Proof of Lemma~\ref{lem3}}: 
\noindent For $W =X,Y$ we have 
\begin{equation}
\label{equatione1}
\EE[W] = \EE[W|\Lambda \in I_0]\PP(\Lambda \in I_0) + \EE[W|\Lambda \in I_1]\PP(\Lambda \in I_1).
\end{equation}
In particular, for $W = X$ we have: $\EE[X] \geq  \EE[X|\Lambda \in
I_0]\PP(\Lambda \in I_0)$. Note that
(\ref{eqin2}) implies that $\EE[X|\Lambda \in I_0] \geq B \cdot
\EE[Y|\Lambda \in I_0]$, so
\begin{equation}
\label{equatione2}
\EE[X] \geq  B \cdot \EE[Y|\Lambda \in I_0]\PP(\Lambda \in I_0).
\end{equation}
Taking $W=Y$ in (\ref{equatione1}) and applying (\ref{eqin}) gives us $$\EE[Y] \leq \EE[Y|\Lambda \in I_0](\PP(\Lambda \in I_0)+\PP(\Lambda \in I_1))=\EE[Y|\Lambda \in I_0]$$
which combined with (\ref{equatione2}) gives the result. 
\hfill$\Box$

\noindent{\em Proof of Claim (i)}, $\EE_1[Y|P_0 \in I_0] \geq
\EE_1[Y|P_0 \in I_1]$:

\noindent 
We will first bound $\EE_1[Y|P_0 \in I_1]$ above.
Let $\mu(n) = (q_0^{q_0}q_1^{q_1}q_2^{q_2}q_3^{q_3})^n$.  
Now,  conditional on ${\bf n}$ satisfying $F_c$ we have
$$n^{-1} \log\left(\mu(n) /Y(t_0, t_1)\right) = d_{KL}(q,p) + o(1),$$
where $p = (p_0, p_1, p_2, p_3)$ and $q = (q_0, q_1, q_2, q_3)$, and
$d_{KL}$ denotes Kullback-Leibler distance.
Now, $d_{KL}(q,p) \geq \frac{1}{2} \|q-p\|_1^2 \geq \frac{1}{2} |q_0 -
p_0|^2$ (the first inequality is a standard one in probability theory).
In particular, if $p_0 \in I_1$, then $|q_0 - p_0| > s>0$. 
Moreover, $$\EE_1[Y|P_0 \in I_1] \leq \max\{Y(t_0, t_1): p_0(t_0, t_1) \in I_1\}.$$ 
Summarizing,
\begin{equation}
\label{eqo1}
\EE_1[Y|P_0 \in I_1] \leq \max\{Y(t_0, t_1): p_0(t_0, t_1) \in I_1\} < \mu(n)e^{-\frac{1}{2}s^2n + o(n)}.
\end{equation}
In the reverse direction, we have:
$$\EE_1[Y|P_0 \in I_0] \geq A(n)B(n)$$
where 
$$A(n) = \min \left\{Y(t_0, t_1): (t_0, t_1) \in [0, n^{-1}] \times
  [t_1^0, t_1^0+n^{-1}] \right\}$$
and 
$$B(n)= \PP\left((t_0, t_1 \right)
\in [0, n^{-1}] \times [t_1^0, t_1^0 + n^{-1}]).$$
Now,
$$A(n)/\mu(n) =
\left(\frac{p_0^{q_0}p_1^{q_1}p_2^{2q_1}}{q_0^{q_0}q_1^{3q_1}}\right)^n \cdot
(p_1^{\Delta_2-\frac{1}{3}\Delta_0}p_2^{\Delta_1+\Delta_3-\frac{2}{3}\Delta_0})
^{\sqrt{n}}.$$
Now, the first term of this product converges to a constant as $n$ grows
(because $p_0 -q_0, p_1-q_1$ and $p_2-q_1$ 
are each of order $n^{-1}$)  while the condition $F_c$ ensures that the second
term decays no faster than $e^{-C_1 \sqrt{n}}$ for a constant $C_1$. Thus, 
$A(n) \geq C_2\mu(n)e^{-C_1\sqrt{n}}$ for a positive constant $C_2$.
The term $B(n)$ is asymptotically proportional to $n^{-2}$.  Summarizing, for a constant $C_3>0$ (dependent just on $t_1^0$)
$$\EE_1[Y|P_0 \in I_0] \geq  C_3 \mu(n)n^{-2}e^{-C_1\sqrt{n}},$$ which
combined with (\ref{eqo1}) establishes claim (i) for $n$ sufficiently
large. 
\hfill$\Box$

In order to prove claim (ii)  we need some preliminary results. 
\begin{lemma}
Let $\eta<1$.  Then for each $x >0$ there exists a value $K = K(x) < \infty$ that depends continuously on $x$ so that the following holds. For any continuous random variable $Z$ on $[0,1]$ with a smooth density function $f$ that satisfies $f(1) \neq 0$ and
$|f'(z)|<B$ for all $z \in (\eta, 1]$, we have
$$ k \cdot \frac{(\EE[Z^k] -
  \EE[Z^{k+1}])}{\EE[Z^k]} \geq \frac{1}{2}$$
  for all $k \geq K(\frac{B}{f(1)})$. 
\label{lem1}
\end{lemma}
{\em Proof.}
Let $t_k = 1- \frac{1}{\sqrt{k}}$. Then $$\EE[Z^k] = \int_{0}^{t_k} t^k f(t)dt + \int_{t_k}^1 t^k f(t)dt.$$
Now,  $$0 \leq \int_{0}^{t_k} t^k f(t)dt \leq t_k^k \sim
e^{-\sqrt{k}-1/2},$$
where $\sim$ denotes asymptotic equivalence (i.e. $f(k) \sim g(k)$ iff
$\lim_{k \rightarrow \infty} f(k)/g(k) =1$). 
Using integration by parts, 
$$\int_{t_k}^1 t^k f(t)dt =  \frac{1}{k+1}\left.t^{k+1}f(t)\right|_{t_k}^1 -
\frac{1}{k+1}\int_{t_k}^1t^{k+1}f'(t)dt.$$
Now, provided $k>(1-\eta)^{-2}$ we have $t_k>\eta$ and so the absolute value of the second term on the right is at most
$\frac{B}{k+1}\int_{t_k}^1t^{k+1}dt  = \frac{B}{(k+1)(k+2)}(1-t_{k}^{k+2}).$
Consequently, $\left|\EE[Z^k] - \frac{f(1)}{k+1}\right|$ is bounded above by $B$
times a term of order $k^{-2}$.
A similar argument, again using integration by parts, shows that
$\left|k(\EE[Z^k]- \EE[Z^{k+1}]) -  \frac{f(1)}{k+1}\right|$ is bounded above by $B$
 times a term of order $k^{-2}$, and the lemma now follows by some routine analysis.
\hfill$\Box$

\begin{lemma}
\label{lem2}
Let $y = (1+2x)(1-x)^2$. Then for $x \in [0,1)$ and $m \geq 3$ we have
$$\left(\frac{1+2x}{1-x}\right)^m \geq  m^2(1-y).$$
\end{lemma}
{\em Proof.}
$$\left(\frac{1+2x}{1-x}\right)^m = \left(1+ \frac{3x}{1-x}\right)^m
\geq \frac{m(m-1)}{2}\left(\frac{3x}{1-x}\right)^2 \geq \frac{9m(m-1)x^2}{2},$$ and
$m^2(1-y) = m^2(3x^2-2x^3) \leq 3m^2x^2$ and for $m\geq 3$ this upper bound is less than the lower bound in the previous 
expression.
\hfill$\Box$

\noindent {\em Proof of Claim (ii)}, for all $p_0 \in I_0$,  $\frac{\EE_1[X|P_0=p_0]}{\EE_1[Y|P_0=p_0]} \geq 6c^2$: 

Write $\EE_1[W|p_0]$ as shorthand for $\EE[W|P_0=p_0]$.
 Note that, for any $r,s>0$,
$\EE_1[P_0^{n_0}P_1^{r}P_2^{s}|p_0] =
p_0^{n_0}\EE_1[P_1^{r}P_2^{s}|p_0]$. Consequently, if we let 
 $k = k(n) = \frac{1}{3}(n-n_0)$ then, by definition of the $\Delta_i$'s,
\begin{equation}
\label{efrac}
\frac{\EE_1[X|p_0]}{\EE_1[Y|p_0]} = \frac{\EE_1[(P_1P_2^2)^{k}\cdot (P_1^{\Delta_1}P_2^{\Delta_2+\Delta_3})^{\sqrt{n}}|p_0]}{\EE_1[(P_1P_2^2)^{k} \cdot (P_1^{\Delta_2}P_2^{\Delta_1+\Delta_3})^{\sqrt{n}}|p_0]}.
\end{equation}
Now, conditional on ${\bf n}$ satisfying $F_c$ (and since $P_1 \geq P_2$) the following two inequalities hold 
$$P_1^{\Delta_1}P_2^{\Delta_2+\Delta_3} 
= \left(\frac{P_1}{P_2}\right)^{\Delta_1} 
\geq \left(\frac{P_1}{P_2}\right)^{2c} 
\mbox{ and }  
P_1^{\Delta_2}P_2^{\Delta_1+\Delta_3} 
= \left(\frac{P_1}{P_2}\right)^{\Delta_2} 
\leq 1.$$
Applying this to
(\ref{efrac}) gives:
\begin{equation}
\label{efrac2}
\frac{\EE_1[X|p_0]}{\EE_1[Y|p_0]} \geq
\frac{\EE_1\left[(P_1P_2^2)^{k}\cdot
\left(\left.\frac{P_1}{P_2}\right)^{2c\sqrt{n}}\right|p_0\right]}{\EE_1[(P_1P_2^2)^{k}|p_0]}.
\end{equation}

Let $U = \frac{P_1-P_2}{1-P_0}$, 
which takes values between $0$ and $1$ because $P_1 \geq P_2$. 
Since $P_1 + 2P_2 = 1-P_0$, we can write
$P_1 = \frac{1}{3}(1+2U)(1-P_0)$ and $P_2 = \frac{1}{3}(1-U)(1-P_0)$.
Thus,
$P_1P_2^2= \frac{1}{27} (1+2U)(1-U)^2(1-P_0)^3$ and 
$\frac{P_1}{P_2} = \frac{(1+2U)}{(1-U)}$. Substituting these into (\ref{efrac2}), letting $Z = (1+2U)(1-U)^2$ and noting that $\sqrt{n} \geq \sqrt{3k}$ gives
$$\frac{\EE_1[X|p_0]}{\EE_1[Y|p_0]}  \geq 
\frac{\EE_1\left[\left.Z^{k}\cdot
(\frac{1+2U}{1-U})^{2c\sqrt{3k}}\right|p_0\right]}{\EE_1[Z^{k}|p_0]}.$$
Thus, by Lemma~\ref{lem2}, (taking $x=U, y=Z, m= 2c\sqrt{3k})$ we
obtain, for $m \geq 3$,
\begin{equation}
\label{boundx}
\frac{\EE_1[X|p_0]}{\EE_1[Y|p_0]} \geq 12c^2k\cdot
\frac{\left(\EE_1[Z^{k}|p_0]-\EE_1[Z^{k+1}|p_0]\right)}{\EE_1[Z^{k}|p_0]}.
\end{equation}
Now the mapping $(t_0, t_1) \mapsto (P_0, Z)$ is a smooth invertible
mapping between $(0, \infty)^2$ and its image within $(\frac{1}{4}, 1) \times (0,1)$.
Notice that $Z$ approaches $1$ whenever $P_0$ approaches $\frac{1}{4}$ or
$1$ (in particular, even if $t_0, t_1$ are independent, $P_0$ and
$Z$ generally will not be). However over the interval $I_0$ 
the conditional density
$f(Z|P_0=p_0)$ of $Z$ given a value $p_0$ for $P_0$ is smooth and bounded
away from $0$, and its first derivative is also bounded above over
this interval. Consequently, we may apply Lemma~\ref{lem1} (noting that 
the condition that  ${\bf n}$ satisfies $F_c$ ensures 
that $k(n) \geq \frac{1}{4}n -o(n)$) to show that for $n$ sufficiently
large the following inequality holds for all  $p_0 \in I_0$, 
$$k\cdot \frac{\left(\EE_1[Z^{k}|p_0]-\EE_1[Z^{k+1}|p_0]\right)}{\EE_1[Z^{k}|p_0]} \geq 
\frac{1}{2}.$$ 
Applying this to (\ref{boundx}) gives
$\frac{\EE_1[X|p_0]}{\EE_1[Y|p_0]} \geq 6c^2$ as claimed.
 This completes the proof of Claim (ii).

\end{document}

%% file: star.pstex_t
\begin{picture}(0,0)%
\epsfig{file=star.eps}%
\end{picture}%
\setlength{\unitlength}{4144sp}%
\begingroup\makeatletter\ifx\SetFigFont\undefined%
\gdef\SetFigFont#1#2#3#4#5{%
  \reset@font\fontsize{#1}{#2pt}%
  \fontfamily{#3}\fontseries{#4}\fontshape{#5}%
  \selectfont}%
\fi\endgroup%
\begin{picture}(3301,972)(110,-211)
\put(1246,-211){\makebox(0,0)[lb]{\smash{\SetFigFont{8}{9.6}{\rmdefault}{\mddefault}{\updefault}\special{ps: gsave 0 0 0 setrgbcolor}$T_2$\special{ps: grestore}}}}
\put(346,-211){\makebox(0,0)[lb]{\smash{\SetFigFont{8}{9.6}{\rmdefault}{\mddefault}{\updefault}\special{ps: gsave 0 0 0 setrgbcolor}$T_1$\special{ps: grestore}}}}
\put(2150,-211){\makebox(0,0)[lb]{\smash{\SetFigFont{8}{9.6}{\rmdefault}{\mddefault}{\updefault}\special{ps: gsave 0 0 0 setrgbcolor}$T_3$\special{ps: grestore}}}}
\put(3050,-211){\makebox(0,0)[lb]{\smash{\SetFigFont{8}{9.6}{\rmdefault}{\mddefault}{\updefault}\special{ps: gsave 0 0 0 setrgbcolor}$T_0$\special{ps: grestore}}}}
\put(117,299){\makebox(0,0)[lb]{\smash{\SetFigFont{5}{6.0}{\rmdefault}{\mddefault}{\updefault}\special{ps: gsave 0 0 0 setrgbcolor}$t_1$\special{ps: grestore}}}}
\put(1018,303){\makebox(0,0)[lb]{\smash{\SetFigFont{5}{6.0}{\rmdefault}{\mddefault}{\updefault}\special{ps: gsave 0 0 0 setrgbcolor}$t_1$\special{ps: grestore}}}}
\put(1915,304){\makebox(0,0)[lb]{\smash{\SetFigFont{5}{6.0}{\rmdefault}{\mddefault}{\updefault}\special{ps: gsave 0 0 0 setrgbcolor}$t_1$\special{ps: grestore}}}}
\put(1128,569){\makebox(0,0)[lb]{\smash{\SetFigFont{5}{6.0}{\rmdefault}{\mddefault}{\updefault}\special{ps: gsave 0 0 0 setrgbcolor}$t_0$\special{ps: grestore}}}}
\put(230,564){\makebox(0,0)[lb]{\smash{\SetFigFont{5}{6.0}{\rmdefault}{\mddefault}{\updefault}\special{ps: gsave 0 0 0 setrgbcolor}$t_0$\special{ps: grestore}}}}
\put(2034,569){\makebox(0,0)[lb]{\smash{\SetFigFont{5}{6.0}{\rmdefault}{\mddefault}{\updefault}\special{ps: gsave 0 0 0 setrgbcolor}$t_0$\special{ps: grestore}}}}
\put(2860,407){\makebox(0,0)[lb]{\smash{\SetFigFont{5}{6.0}{\rmdefault}{\mddefault}{\updefault}\special{ps: gsave 0 0 0 setrgbcolor}$t_1^0$\special{ps: grestore}}}}
\put(1917, 31){\makebox(0,0)[lb]{\smash{\SetFigFont{6}{7.2}{\rmdefault}{\mddefault}{\updefault}\special{ps: gsave 0 0 0 setrgbcolor}3\special{ps: grestore}}}}
\put(2493, 28){\makebox(0,0)[lb]{\smash{\SetFigFont{6}{7.2}{\rmdefault}{\mddefault}{\updefault}\special{ps: gsave 0 0 0 setrgbcolor}2\special{ps: grestore}}}}
\put(2219, 28){\makebox(0,0)[lb]{\smash{\SetFigFont{6}{7.2}{\rmdefault}{\mddefault}{\updefault}\special{ps: gsave 0 0 0 setrgbcolor}1\special{ps: grestore}}}}
\put(2814, 28){\makebox(0,0)[lb]{\smash{\SetFigFont{6}{7.2}{\rmdefault}{\mddefault}{\updefault}\special{ps: gsave 0 0 0 setrgbcolor}1\special{ps: grestore}}}}
\put(3110, 28){\makebox(0,0)[lb]{\smash{\SetFigFont{6}{7.2}{\rmdefault}{\mddefault}{\updefault}\special{ps: gsave 0 0 0 setrgbcolor}2\special{ps: grestore}}}}
\put(3392, 31){\makebox(0,0)[lb]{\smash{\SetFigFont{6}{7.2}{\rmdefault}{\mddefault}{\updefault}\special{ps: gsave 0 0 0 setrgbcolor}3\special{ps: grestore}}}}
\put(1010, 29){\makebox(0,0)[lb]{\smash{\SetFigFont{6}{7.2}{\rmdefault}{\mddefault}{\updefault}\special{ps: gsave 0 0 0 setrgbcolor}2\special{ps: grestore}}}}
\put(1585, 27){\makebox(0,0)[lb]{\smash{\SetFigFont{6}{7.2}{\rmdefault}{\mddefault}{\updefault}\special{ps: gsave 0 0 0 setrgbcolor}1\special{ps: grestore}}}}
\put(110, 28){\makebox(0,0)[lb]{\smash{\SetFigFont{6}{7.2}{\rmdefault}{\mddefault}{\updefault}\special{ps: gsave 0 0 0 setrgbcolor}1\special{ps: grestore}}}}
\put(419, 28){\makebox(0,0)[lb]{\smash{\SetFigFont{6}{7.2}{\rmdefault}{\mddefault}{\updefault}\special{ps: gsave 0 0 0 setrgbcolor}2\special{ps: grestore}}}}
\put(686, 28){\makebox(0,0)[lb]{\smash{\SetFigFont{6}{7.2}{\rmdefault}{\mddefault}{\updefault}\special{ps: gsave 0 0 0 setrgbcolor}3\special{ps: grestore}}}}
\put(1320, 28){\makebox(0,0)[lb]{\smash{\SetFigFont{6}{7.2}{\rmdefault}{\mddefault}{\updefault}\special{ps: gsave 0 0 0 setrgbcolor}3\special{ps: grestore}}}}
\end{picture}